\documentclass[10pt]{article}
\usepackage[utf8]{inputenc}
\usepackage[T1]{fontenc}
\usepackage{amsmath}
\usepackage{amsthm}
\usepackage{amsfonts}
\usepackage{amssymb}
\usepackage{makeidx}
\usepackage{graphicx}
\usepackage{lmodern}
\usepackage{mathtools}
\usepackage{url}
\usepackage{xcolor}
\usepackage[left=2cm,right=2cm,top=1cm,bottom=2cm]{geometry}
\usepackage{multirow}
\usepackage{caption}
\usepackage{subcaption}
\usepackage{tabularx} 
\usepackage{booktabs}
\usepackage{algorithm}
\usepackage{algorithmic}

\author{Fermat Leukam     \thanks{AIMS South Africa /Stellenbosch university, South Africa. }
 \and   Rock Stephane Koffi\thanks{ University of South Africa, South Africa}
  \and   Prudence Djagba\thanks{ Michigan State University, USA}
  }
\title{Reinforcement Learning for Portfolio Optimization with a Financial Goal and Defined Time Horizons}

\usepackage{hyperref}
\usepackage{cleveref}

\theoremstyle{remark}

\theoremstyle{definition}

\date{}
\begin{document}
\maketitle	
\begin{abstract} 
\noindent

This research proposes an enhancement to the innovative portfolio optimization approach using the G-Learning algorithm, combined with parametric optimization via the GIRL algorithm (G-learning approach to the setting of Inverse Reinforcement Learning) as presented by \cite{dixon2020machine}. The goal is to maximize portfolio value by a target date while minimizing the investor's periodic contributions. Our model operates in a highly volatile market with a well-diversified portfolio, ensuring a low-risk level for the investor, and leverages reinforcement learning to dynamically adjust portfolio positions over time. Results show that we improved the Sharpe Ratio from 0.42, as suggested by recent studies using the same approach, to a value of 0.483 a notable achievement in highly volatile markets with diversified portfolios. The comparison between G-Learning and GIRL reveals that while GIRL optimizes the reward function parameters (e.g., \( \lambda = 0.0012 \) compared to 0.002), its impact on portfolio performance remains marginal. This suggests that reinforcement learning methods, like G-Learning, already enable robust optimization. This research contributes to the growing development of reinforcement learning applications in financial decision-making, demonstrating that probabilistic learning algorithms can effectively align portfolio management strategies with investor needs.
 \\


\end{abstract}
\section{Introduction}

Portfolio optimization is a central challenge in finance, with the primary goal of identifying the optimal combination of assets to maximize returns for a given level of risk or minimize risk for a desired level of return. One of the earliest approaches proposed to address this challenge is the Modern Portfolio Theory (MPT), introduced by  \cite{markowitz1952modern}. Today, with the advent of artificial intelligence, increasingly effective solutions are emerging. The variant of this problem, which involves optimizing a portfolio to achieve a specific goal, has also gained attention. This problem is essential for both private and institutional investors as it determines their ability to meet specific financial objectives within a given time horizon.

Throughout our lives, we are required to meet financial goals at specific dates. Holding a portfolio that allows us to achieve these goals in a constantly changing financial environment characterized by unpredictable events such as economic crises or pandemics, without excessive effort, is highly relevant, both for individual investors and large organizations. Such disruptions highlight the need for dynamic and adaptive strategies. Therefore, efficient portfolio management is crucial, not only for wealth preservation but also for achieving long-term objectives, such as retirement planning or acquiring major assets.

The importance of this research lies in the need to improve traditional portfolio optimization methods and enhance the effectiveness of supervised learning models. Despite their influence, these methods present several limitations. In particular, the assumptions of normally distributed returns and independent variables often prove inadequate in real-world financial markets, which are marked by extreme volatility and unforeseen shocks. The evolution of reinforcement learning offers more flexible and effective solutions that can adapt to the complex dynamics of financial markets.

The work of Markowitz laid the foundation for portfolio management by introducing the concept of the \textit{efficient frontier} \cite{markowitz1952modern}. However, scholars like Benoît Mandelbrot and Nassim Nicholas Taleb have shown that financial markets often follow power laws, limiting the application of Markowitz's approach \cite{hubbard2020failure}. The rise of artificial intelligence has enabled the development of new solutions, with many relying on supervised learning methods \cite{CFAinstitut}. However, a major limitation of these methods is their inability to adapt to external events that impact markets, such as financial crises or pandemics. The evolution of reinforcement learning offers more promising solutions to these challenges. Specifically, direct reinforcement algorithms such as Deep
Deterministic Policy Gradient (DDPG), Soft Actor-Critic (SAC), Proximal Policy Optimization (PPO), Actor-Critic(A2C), and Twin Delayed Deep Deterministic Policy Gradient(TD3) have shown encouraging results \cite{hachaichi2024benchmarking}. Recent studies, such as those by \cite{jiang2017deep}, have introduced deep reinforcement learning (DRL), demonstrating its effectiveness in managing dynamic portfolios.

However, most existing research focuses on the formulation of classical problems, maximizing returns or minimizing risk, without considering investor-specific objectives at a given date \cite{jin2016portfolio, jiang2017deep}. Goal-based wealth management has emerged as a more relevant approach, as demonstrated by the work of \cite{browne2000stochastic} and \cite{das2020dynamic}. This approach leverages Markov Decision Processes (MDP) to maximize the probability that the final portfolio value will exceed a target amount. G-Learning has already been applied to dynamic portfolio optimization in \cite{halperin2018market}, with further interesting extensions for problems involving cash flows developed in \cite{dixon2020g}.

This research aims to develop a dynamic portfolio optimization solution using G-Learning, a probabilistic extension of Q-learning. Unlike traditional and supervised learning approaches, our method takes into account both the evolution of regular contributions and the achievement of a specific goal by a given date. It also adapts to the high volatility and dynamic nature of financial markets. We apply this approach to a practical case: optimizing a portfolio intended to fund the purchase of a vehicle by a specific date, with regular contributions made over time.

The approach consists of:
\begin{itemize}
    \item Maximizing the portfolio value at the target date.
    \item Minimizing the investor's regular contributions throughout the investment period.
\end{itemize}

We hypothesize that the use of G-Learning can produce a more efficient investment strategy than traditional and supervised learning methods by maximizing returns and minimizing the financial effort required to achieve a goal by a specific date, with low-risk portfolios.

The structure of this report is as follows: In the first section, we will discuss the fundamental concepts related to portfolio optimization and reinforcement learning. Next, we will outline the approach for regularizing Q-learning and the development of the G-Learner and GIRL algorithms, followed by a data simulation. This will be followed by a presentation of the results obtained. Finally, we will provide a discussion and conclusion, along with some future perspectives.

\section{Background} 

In this chapter, we will present the fundamental knowledge necessary for understanding the problem we aim to solve and the approach we will use.

\subsection{Portfolio optimization}
Modern Portfolio Theory (MPT) \cite{markowitz1952modern}, also known as mean-variance analysis, provides a mathematical approach to constructing a portfolio that aims to maximize returns for a given amount of risk. It builds on the principle of diversification, which suggests that holding a variety of asset types reduces overall risk compared to concentrating investments in a single category \cite{cvitanic2001theory}. A core insight of MPT is that the value of an asset is not in its individual risk and return, but in how it influences the overall performance of the portfolio. This theory was first introduced by Harry Markowitz in his 1952 doctoral thesis, where he developed what is now known as the Markowitz mode. 

Portfolio optimization is the process of selecting a combination of assets and determining their respective weights in a way that aligns with the investor's objectives. This involves balancing the trade-off between risk and return to build a portfolio that maximizes expected returns while minimizing financial risks and other costs. As a multi-objective optimization problem, it seeks to identify the most efficient allocation of assets to achieve the desired performance \cite{milhomem2020analysis}.

In the following lines, we define a set of financial concepts that are essential for understanding and solving the portfolio optimization problem.

\subsubsection{Asset}
An asset refers to any resource with economic value that a company, individual, or nation controls or owns, with the expectation that it will generate future benefits.

In numerous portfolio optimization scenarios, the focus is on financial instruments, such as equities, corporate and government bonds, along with various other forms of securities. A stock represents ownership in a company, with each unit referred to as a share. In contrast, bonds are generally considered lower-risk investments with more modest returns compared to stocks, making them a key component in diversified portfolios, particularly for conservative or older investors seeking stability.

\subsubsection{Returns}
Before defining the return of a portfolio, it is necessary to understand what the return of an asset is.

\textbf{Return} or \textbf{Realized Return} refers to the profit or loss produced by an investment over a specific period, typically represented as a percentage of the original amount invested.

Here is the formula for calculating Return on Investment (ROI):
\[
\text{ (ROI)} = \frac{\text{Gain from Investment} - \text{Cost of Investment}}{\text{Cost of Investment}}.
\]

\textbf{Expected Return} refers to the anticipated return on an investment based on historical data, statistical models, or forecasts. It represents the average return that an investor expects or estimates will be achieved in the future, considering the probability of various possible outcomes.

\[
\text{Expected Return} = \sum_{i=1}^{n} p_i \times r_i.
\]

where:
\begin{itemize}
    \item \( p_i \) is the probability of \( i^{th} \) outcome. 
    \item \( r_i \)  denotes the return associated with the \( i^{th} \) outcome.
    \item \( n \) indicates the total number of potential outcomes.
\end{itemize}

 The \textbf{Portfolio Return} (\( R_d \)) is the total return generated by a portfolio, determined by taking the weighted average of the returns of its individual assets. It reflects the overall performance an investor can anticipate from the combined holdings.

\[
R_d := \sum_{a=1}^m w_a R_a.
\]

where:  
\begin{itemize}  
    \item \( R_d \) represents the overall return of the portfolio.  
    \item \( w_a \) signifies the proportion of the \(a^{th}\) asset within the portfolio.  
    \item \( R_a \) indicates the return of the \(a^{th}\) asset.  
    \item \( m \) denotes the total count of assets included in the portfolio.  
\end{itemize}

\subsubsection{Risk}
A widely accepted international definition of risk is the "impact of uncertainty on objectives." The interpretation of risk, along with the techniques for assessment and management, as well as its descriptions and definitions, can vary across different domains. In the financial sector, which is our focus, risk refers to the likelihood that the actual return on an investment will differ from what was anticipated. This encompasses not only "downside risk" (returns falling short of expectations, which may include the potential loss of part or all of the initial investment) but also "upside risk" (returns that exceed initial expectations) \cite{sethi2013survey}.

In finance, risk is commonly quantified using variance and standard deviation of returns. These measures help investors understand the volatility or variability in the returns of an investment or portfolio.

The \textbf{variance of the portfolio return} (\( \sigma_p^2 \)) quantifies the spread of returns across the entire portfolio. This measure takes into account the variance of each individual asset as well as the covariance among various assets within the portfolio.

\[
\sigma_p^2 = \sum_{i} w_i^2 \sigma_i^2 + \sum_{i} \sum_{j \neq i} w_i w_j \sigma_i \sigma_j \rho_{ij}.
\]

where:  
\begin{itemize}  
    \item \( \sigma_p^2 \) represents the variance of the portfolio return.  
    \item \( w_i \) signifies the proportion of asset \(i\) within the portfolio.  
    \item \( \sigma_i^2 \) denotes the variance of the return associated with asset \(i\).  
    \item \( \sigma_i \) indicates the standard deviation of the return for asset \(i\).  
    \item \( \rho_{ij} \) represents the correlation coefficient between the returns of assets \(i\) and \(j\).  
    \item The first term (\( \sum_{i} w_i^2 \sigma_i^2 \)) reflects the variance of individual assets.  
    \item The second term (\( \sum_{i} \sum_{j \neq i} w_i w_j \sigma_i \sigma_j \rho_{ij} \)) accounts for the covariance among different assets.  
\end{itemize}

The \textbf{standard deviation of the portfolio return} (\( \sigma_p \)) measures the portfolio's overall volatility. It is simply the square root of the variance of the portfolio return.

\subsubsection{Diversification}
Diversification is a risk management approach that involves combining different investments within a portfolio. A well-diversified portfolio includes a variety of asset classes and investment options to reduce reliance on any single asset or risk factor. The underlying principle of this strategy is that a portfolio made up of diverse assets will generally provide higher long-term returns while minimizing the risk associated with individual holdings or securities \cite{manganelli2010finance}.

\subsubsection{Sharpe Ratio}
In finance, the \textbf{Sharpe Ratio} is a metric utilised to assess the performance of an investment, such as a securities or portfolio, by contrasting it with a risk-free asset, subsequent to risk adjustment. The Sharpe Ratio is calculated by subtracting the risk-free return from the investment returns and dividing the result by the standard deviation of the investment returns. This ratio signifies the excess return an investor obtains for the increased volatility associated with holding a riskier asset \cite{gatfaoui2015estimating}.

\[
\text{Sharpe Ratio} = \frac{E[R_a - R_f]}{\sigma_a}.
\]

where:  
\begin{itemize}  
    \item \( R_a \) denotes the return on the asset.  
    \item \( R_f \) represents the return from a risk-free investment.  
    \item \( E[R_a - R_f] \) signifies the expected value of the asset's excess return above the risk-free return.  
    \item \( \sigma_a \) indicates the standard deviation of the asset's excess return.  
\end{itemize}

\subsubsection{Financial markets}
A financial market is a venue or system dedicated to the exchange of assets \cite{scott1976teaching}. These markets play an essential role in allocating funds to economic activities with the goal of maximizing returns. They encompass various trading environments such as the stock market, bond market, foreign exchange market, and derivatives market. Therefore, financial markets are fundamental to the smooth operation of capitalist economies.

These markets create financial products that generate returns for investors or lenders with surplus funds while providing borrowers access to the capital they need to finance their projects.

\subsubsection{Efficient frontier}
\begin{figure}[H]
    \centering
    \includegraphics[width=0.7\linewidth]{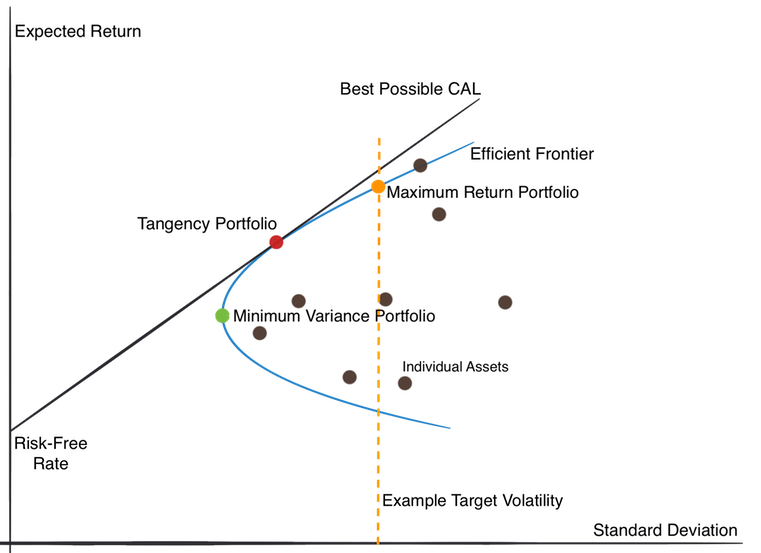}
    \caption{Visualization of the concept of the Efficient Frontier.}
    \label{fig:Efficient_Frontier}
\end{figure}

Mathematical approaches to portfolio optimization rely on the concept of the Efficient Frontier ~Fig. \ref{fig:Efficient_Frontier} \cite{beste2002markowitz}, which is a hyperbolic curve representing the set of optimal portfolios formed from various combinations of assets. These portfolios are defined by two key principles: they either aim to minimize risk for a specific target return or seek to maximize returns for a predetermined level of risk \cite{DissUMMTO}. Optimizing the portfolio within this framework, therefore, consists of selecting the portfolio on this frontier that meets the desired requirements.

In approaches based on reinforcement learning, the goal is to define a reference portfolio. In our case, we will use the Benchmark portfolio.

\subsubsection{Benchmark portfolio}
A benchmark portfolio is a model portfolio used as a point of comparison to evaluate the performance of an actual portfolio \cite{maleyeff2003benchmarking}. It serves several essential functions:
\begin{itemize}
    \item Passive exposure: It acts as a template for investors seeking to passively track the performance of a specific market segment.
    \item Performance evaluation: It measures the added value generated by active managers by comparing their results with those of the benchmark portfolio.
\end{itemize}

\subsection{Reinforcement Learning}
The first concepts that come to mind when discussing machine learning are supervised learning and unsupervised learning. Let’s first clarify the difference between these types of learning and reinforcement learning.

\textbf{Supervised learning} involves training an agent to produce an output based on a given input, using examples provided by a "teacher" in the form of input-output pairs. The agent's goal is to generalize from these examples, meaning to find a function that generates the correct outputs from the corresponding inputs.

\textbf{Unsupervised learning} refers to a category of machine learning where algorithms are trained using data that lacks explicit labels or defined outcomes. The objective is for the model to autonomously discover hidden patterns, structures, or relationships within the dataset.

\textbf{Reinforcement Learning (RL)} is another form of machine learning in which an agent learns to make decisions through interactions with an environment to maximize a cumulative reward. The agent takes various actions, observes the outcomes, and modifies its behavior based on feedback (rewards or penalties) to enhance its performance over time \cite{dixon2020machine}. From a mathematical perspective, this process can be framed as a problem of maximizing a specific objective function.

To understand the concept of reinforcement learning, it is essential to define the various elements involved in this concept and to present the different versions of this learning approach.

\subsubsection{Markov Decision Processes}
Markov Decision Processes (MDPs) \cite{markowitz1952modern} provide a mathematical framework for representing decision-making scenarios in which outcomes are influenced by both random factors and the actions of an agent. They build upon traditional Markov models by incorporating extra variables to capture the actions or controls implemented by the agent. Specifically, in the realm of reinforcement learning, MDPs are crucial as they establish a structure to model the interactions between an agent and its surrounding environment.

An MDP is characterized by a series of discrete time intervals \( t_0, t_1, \ldots, t_n \) and a tuple \( (S, A, p, R, \gamma) \), where:
\begin{itemize}
    \item \( S \): Represents the collection of all potential states of the environment. At each time interval \( t \), the system resides in a specific state \( S_t \in S \). This collection can either be continuous or discrete.
     \item \( A \): Set of possible actions that the agent can take. The choice of an action depends on the current state \( S_t = s \). This set can also be discrete or continuous.
     \item \( p(s' | s, a) \): These are the transition probabilities, which indicate the likelihood of reaching a new state \( s' \) from the current state \( s \) after performing an action \( a \). This concept outlines how the environment changes in reaction to the actions taken by the agent.
     \item \( R \): Reward function that assigns a value to the transition between states based on the action taken: \( R(s, a, s') \). It indicates the expected reward when the agent moves from \( s \) to \( s' \) after taking action \( a \).
     \item \( \gamma \): This is the discount factor, which is a value ranging from 0 to 1 that assesses the significance of future rewards in comparison to present rewards. A \( \gamma \) value approaching 1 places greater emphasis on future rewards, whereas a value nearing 0 prioritizes immediate rewards.
\end{itemize}
One of the important characteristics of Markov Decision Processes is that the conditional distribution of \(s_{t+1}\), given the past (i.e., knowing \(\left(S_{k}\right)_{0 \leq k \leq t}\), depends only on \(s_t\):
\[
P(S_{t+1} = s_{t+1} \mid S_0 = s_0, S_1 = s_1, \dots, S_t = s_t) = P(S_{t+1} = s_{t+1} \mid S_t = s_t).
\]

The MDP approach provides a formal framework for the reinforcement learning process, modeling the dynamics of the environment and the outcomes of actions. Through algorithms such as Q-learning, MDPs enable the agent to estimate the value of different actions and improve its policy to maximize the expected reward.

\subsubsection{Elements of Reinforcement Learning}
\begin{figure}[H]
    \centering
    \includegraphics[width=0.9\linewidth]{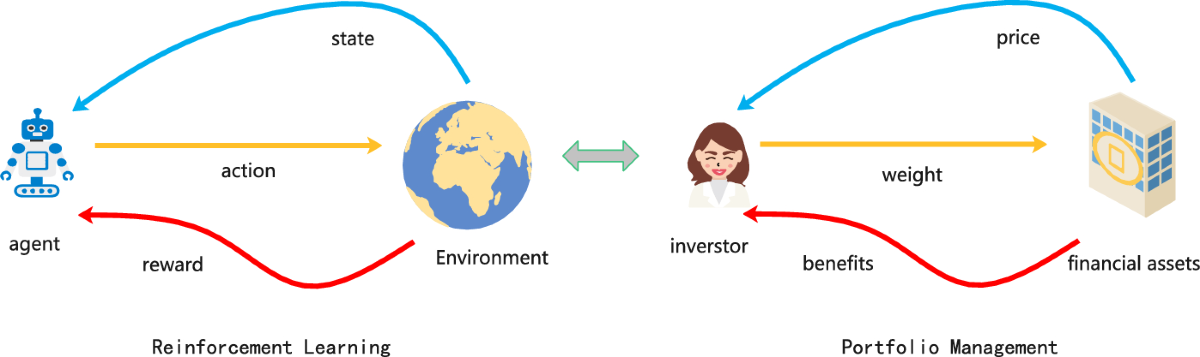}
    \caption{Emphasizing the elements of reinforcement learning for portfolio optimization.}
    \label{fig:RL_Portfolio}
\end{figure}

The use of reinforcement learning to address the portfolio optimization problem requires a specification of the reinforcement learning elements related to this issue, as illustrated in ~Fig. \ref{fig:RL_Portfolio} \cite{li2024deep}.

Therefore, it is essential to define these key elements of reinforcement learning in order to better identify them within the context of our problem.The basic elements of reinforcement learning are: Environment, agent, policy, reward function, value function, and action-value function.
\begin{enumerate}
    \item \textbf{Environment and Agent}
    Reinforcement learning (RL) consists of two primary components: the environment and the agent. \textbf{The environment} refers to the setting in which the agent operates, while \textbf{the agent} is the entity that engages with the environment. RL environments possess several key features, including the state of the environment, the possible actions available to the agent, and the rewards received by the agent following any transition between states. The state encapsulates all relevant information about the environment at a specific moment. As the agent takes actions, the state of the environment changes, leading to a transition and the provision of a reward to the agent.

   \item \textbf{Policy function}
   
 A policy function \(\pi_t(S_t)\) outlines the actions an agent should take at time \(t\) based on the current state \(S_t\) of the environment. This policy can either be deterministic, providing a specific action for each state, or it can represent a probability distribution over the available actions, depending on \(S_t\).

\item \textbf{Reward function}

A \textbf{reward function} in reinforcement learning defines the goal of the agent by assigning a numerical reward signal based on the current state, action, or transition between states. It quantifies the immediate benefit or penalty of taking an action in a specific state, guiding the agent towards achieving its objectives.

Mathematically, the reward function \( R(s, a, s') \) yields a reward \( r \) for the transition from state \( s \) to state \( s' \) following action \( a \). Typically, the reward function can be influenced by the current state \( s \), the action \( a \), and the resulting state \( s' \). The agent aims to maximize its total rewards over time, making the reward function essential for guiding its learning and behavior.

\item \textbf{Value function}

The state-value function \( V^{\pi}_{t}(s) \) represents the anticipated discounted return when beginning from state \( s \), meaning \( S_{t} = s \), and then consistently adhering to policy \( \pi \). In simpler terms, this function assesses the quality or desirability of being in a specific state
\cite{dixon2020machine}.

\[
V^{\pi}_{t}(s) = \mathbb{E}^{\pi}_{t} \left[ \sum_{i=0}^{T-t-1} \gamma^{i} R(S_{t+i}, a_{t+i}, S_{t+i+1}) \mid S_t = s \right].
\]

In this context, \( R(S_{t+i}, a_{t+i}, S_{t+i+1}) \) denotes the reward obtained at time \( t+i \). The variable \( T \) signifies the planning horizon (with \( T = \infty \) indicating an infinite-horizon scenario). Additionally, \( \mathbb{E}^{\pi}_{t}[\cdot \mid S_t = s] \) indicates the conditional expectation across all potential future states, assuming actions are determined by policy \( \pi \). Finally, \( \gamma \) represents the discount factor, constrained within the range \( 0 \leq \gamma \leq 1 \). 

A simple transformation of this expression gives us the recurring formulation below, known as the Bellman equation, which is much more interesting in practice. 
\[
V^\pi_t(s) = \mathbb{E}^{\pi} \left[ R(s, a_t, S_{t+1}) + \gamma V^\pi_{t+1}(S_{t+1}) \; \Big| \; S_t = s \right].
\]
AndThe optimal value function \( V_t^* \) signifies the maximum value for a given state across all potential policies, achieved when following the optimal policy \( \pi^* \) \cite{dixon2020machine}: 
\[
V^*_t(s) := V^{\pi^*}_t(s) = \max_{\pi} V^\pi_t(s), \quad \forall s \in S.
\]

\item \textbf{Action-Value Function.}
The action-value function \( Q(s, a) \) (also known as Q-value) reflects the worth of executing a specific action in a given state. It evaluates the anticipated future rewards achievable by beginning in that state, performing the action, and subsequently adhering to an optimal policy\cite{Reward}.
\[
Q^{\pi}_t(s, a) = \mathbb{E}^{\pi}_t \left[ \sum_{i=0}^{T - t - 1} \gamma^i R\left(S_{t+i}, a_{t+i}, S_{t+i+1}\right) \mid S_t = s, A_t = a \right].
\]
 Its optimal Bellman expression in relation to the value function is given by:
 \[
Q^*_t(s, a) = \mathbb{E}^*_t \left[ R_t(s, a, s') \right] + \gamma \, \mathbb{E}^*_t \left[ V^*_{t+1}(s') \right].
\]
This concept plays a crucial role in reinforcement learning, particularly within Q-learning. It facilitates the creation of the Q-table a lookup table where each entry predicts the total reward accrued by taking a specific action in a particular state and subsequently adhering to the optimal strategy.

\end{enumerate}

\subsubsection{Q-learning}

\begin{figure}[H]
    \centering
    \includegraphics[width=0.9\linewidth]{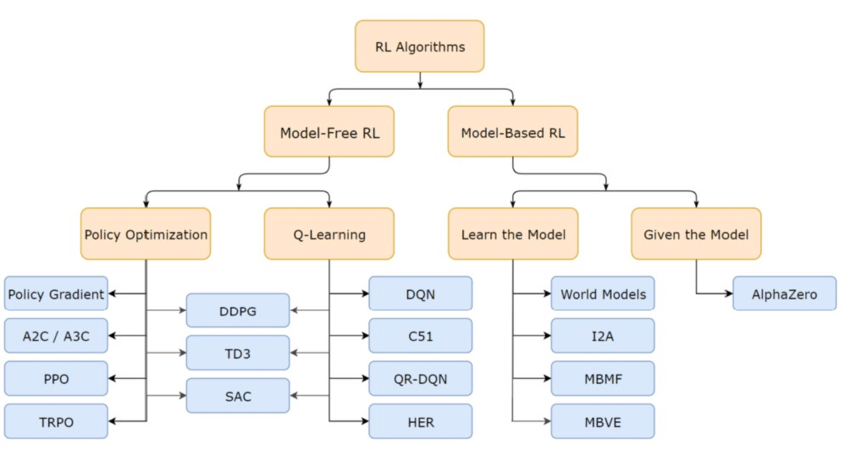}
    \caption{Taxonomy of RL models.}
    \label{fig:Taxonomy}
\end{figure}

Before presenting Q-learning, which will be the focus of our study later, let’s first introduce the taxonomy of reinforcement learning methods ~Fig. \ref{fig:Taxonomy} \cite{salvador2021my}. These methods are divided into two main families.
On one hand, we have Model-Free methods, where the agent learns to make decisions solely based on its direct experiences with the environment. On the other hand, Model-Based methods rely on the agent’s ability to learn or have access to a model of the environment, which can predict state transitions and rewards.
These two families are further divided into subcategories. The first distinction is based on the availability of the model: the agent either learns the model from experience or has the model provided. The second distinction concerns the availability of the policy, with methods focusing either on policy optimization or Q-learning.These different subfamilies include very recent and widely used algorithms in reinforcement learning.

Q-learning is a reinforcement learning technique designed to discover the best action-selection strategy for a finite Markov decision process (MDP). It enables an agent to optimize cumulative rewards by interacting with the environment repeatedly, even in cases where the underlying dynamics of the environment remain unknown \cite{salvador2021my}.This is a very popular algorithm in reinforcement learning, proposed by \cite{watkins1992q}. Before explaining how it works, let's start by recalling the two groups of learning methods, depending on policy. These are the on-policy and off-policy algorithms.

On-policy algorithms operate under the assumption that the policy generating the data for learning is already optimal, aiming to derive the best policy function from that dataset. Conversely, off-policy algorithms do not require the data-generating policy to be optimal; it can even be sub-optimal or random. The goal of off-policy methods is to identify an optimal policy using data collected under a different, potentially non-ideal, policy.

Q-learning is therefore an off-policy reinforcement learning algorithm, so the data is used to update the Q-table in order to determine the policy.Let us recall that the dataset is a set of trajectories comprising tuples \((s_t, a_t, R_t, s_{t+1})\) that have been collected through the observation of an agent in the environment we want to study, or in a similar environment, under a set of various policies.
 It works as follows:
\subsection*{a. Learning and Updating Q-values}
The algorithm stores a table of Q-values corresponding to each state-action pair. These Q-values indicate the anticipated utility of executing a specific action within a particular state, followed by adherence to the optimal policy. Initially, the Q-values are assigned arbitrary values and are progressively refined through iterative updates based on the agent's accumulated experiences.

\subsection*{b. Q-value Update Rule}
The Q-values are refined through the following equation:

\[
Q(s, a) \leftarrow Q(s, a) + \alpha \left[ r + \gamma \max_{a'} Q(s', a') - Q(s, a) \right].
\]

Where:
\begin{itemize}
    \item \(a'\) denotes a potential action from the next state \(s'\).
    \item \(\alpha\) is the learning rate, with \(0 < \alpha \leq 1\).
\end{itemize}

\subsection*{c. Policy Derivation}
The policy defines the action to execute in each state and is guided by the Q-values. In most cases, it selects the action with the highest Q-value for a given state (exploitation). However, to encourage exploration, it may occasionally opt for a suboptimal action.

\subsection*{d. Exploration vs. Exploitation}
Q-learning balances the need to explore new strategies with the goal of exploiting known ones. This trade-off is typically handled through techniques like the epsilon-greedy strategy, where the agent primarily selects the most favorable action based on prior experience but occasionally opts for a random action to uncover alternative solutions.

\subsection*{e. Convergence}
Given specific conditions, such as guaranteeing that every state action pair is explored infinitely, Q-learning will eventually converge toward the optimal policy and the corresponding Q values, yielding the highest expected reward for each state, regardless of the circumstances.

\section{Methods}%

In this chapter, we will present the reinforcement learning method for portfolio optimization proposed by \cite{dixon2020g}, where they develop two algorithms, G-Learner and GIRL, derived from an entropy regularization of  Q-learning method.

\subsection{G-learning: Mathematical Foundations and Algorithms}
G-Learning is an extension of Q-learning, tailored to complex financial environments. Models a sequential decision-making process in which an agent must make choices at each time step to optimize a long-term objective, such as maximizing portfolio wealth.

\subsubsection{Mathematical Approach of G-learning and its Algorithms G-learner and GIRL}
We will demonstrate here, from a mathematical point of view, how G-learning is established through regularization of the Q-learning method. We will also explain the construction of the G-learner and GIRL algorithms.

\subsection*{a. Regularization of Q-learning.}
The objective of Reinforcement Learning (RL) is to address the Bellman optimality equation using data samples.

In the context of Q- learning, we pose \(V^*_t(s) = \max_a Q^*_t(s, a)\) and \(\pi^*_t(a_t \mid x_t) = \arg \max_{a_t}  Q^*_t(x_t, a_t)\)
However, in the context of G-learning, which is a probabilistic extension of Q-learning, it is necessary to regularize the Q function to obtain the expression of the action value function to be used. We will call this function G in what follows.

We begin by rewriting Bellman's optimality equation for the value function.

Let \( P = \{ \pi : \pi \geq 0; \ \mathbf{1}^T \pi = 1 \} \), which is a Fenchel-type representation of \( V^*_t \), we have:

\begin{equation}
    V^*_t(x_t) = \max_{\pi(.|y) \in P} \sum_{a_t \in A_t} \pi(a_t|x_t) \left[ R^*_t(x_t, a_t) + \gamma \mathbb{E}_{t, a_t} \left[ V_{t+1}(x_{t+1}) \right] \right].
    \label{eq:1}
\end{equation}

In fact, this follows from the fact that \( \max_{i \in \{1, \dots, n\}} x_i = \max_{\pi \geq 0; \|\pi\|_1 \leq 1} \pi^T x \).
Now, let us use regularized entropy to construct our G function from this expression. For simplicity, we will represent the expectation. 
\(\mathbb{E}_{x_{t+1} | x_t, a_t}[\cdot]\) 
as \(\mathbb{E}_{t, a}[\cdot]\) in the following.

This new formulation of the value function is more interesting for the rest of our work. We also need to define the following concepts:
\begin{itemize}
    \item\textbf{One-step information cost} measures how much the learned policy \( \pi(a_t | x_t) \) differs from a reference policy \( \pi_0(a_t | x_t) \) for a given state \( x_t \) and action \( a_t \). It is defined as:
   \begin{equation}
       g^\pi(x_t, a_t) := \log \frac{\pi(a_t | x_t)}{\pi_0(a_t | x_t)}  .
       \label{eq:2}
   \end{equation}
  
  \item \textbf{Expected information cost} is the average of the one-step information cost over all possible actions, weighted by the probability of those actions under the learned policy. Mathematically, this is the \textit{Kullback-Leibler (KL) divergence} between the learned policy \( \pi(a_t | x_t) \) and the reference policy \( \pi_0(a_t | x_t) \):
   \begin{equation}
       \mathbb{E}_\pi[g^\pi(x_t, a_t) | x_t] = \text{KL}[\pi || \pi_0](x_t) := \sum_{a_t} \pi(a_t | x_t) \log \frac{\pi(a_t | x_t)}{\pi_0(a_t | x_t)} .
       \label{eq:3}
   \end{equation}
   
  \item \textbf{Total discounted information cost} is the cumulative information cost over a series of actions (trajectory) from time step \( t \) to the terminal time step \( T \). It is discounted by a factor \( \gamma \in [0,1] \), where future costs are valued less than immediate ones:
   \begin{equation}
        I^\pi(x_t) := \sum_{t' = t}^T \gamma^{t' - t} \mathbb{E}^\pi_t \left[ g^\pi(x_{t'}, a_{t'}) | x_t \right] .
        \label{eq:4}
   \end{equation}

  \item \textbf{Free energy function} combines the standard value function \( V_t^\pi(x_t) \) with a penalty for the total discounted information cost \( I^\pi(x_t) \), weighted by a regularization parameter \( \frac{1}{\beta} \). This regularization ensures that the learned policy remains close to the reference policy while optimizing rewards. The free energy function is defined as:
  \begin{equation}
       F_t^\pi(x_t) := V_t^\pi(x_t) - \frac{1}{\beta} I^\pi(x_t) .
       \label{eq:5}
   \end{equation}
   
  Using the expression for \(V_t^\pi(x_t)\) and relation Eq.~\eqref{eq:5}
this can be expanded as:
   \begin{equation}
       F_t^\pi(x_t) = \sum_{t' = t}^T \gamma^{t' - t} \mathbb{E}_\pi \left[ R_{t'}(x_{t'}, a_{t'}) - \frac{1}{\beta} g^\pi(x_{t'}, a_{t'}) \right] . 
       \label{eq:6}
   \end{equation}
  
   Here, \( \beta \) controls the trade-off between optimizing rewards (through \( V_t^\pi(x_t) \)) and minimizing the divergence from the reference policy (through \( I^\pi(x_t) \)).
\end{itemize}

The free energy \(F_t^\pi(x_t)\) is thus the value function regularized by entropy. It replaces the value function in the G-learning approach

 Its Bellman equation is given by :

\begin{equation}
    F_t^\pi(x_t) = \mathbb{E}_{a_t | x_t} \left[\hat{R}_t(x_t, a_t) - \frac{1}{\beta} g^\pi(x_t, a_t) + \gamma \mathbb{E}_{t, a} \left[ F_{t+1}^\pi(x_{t+1}) \right] \right] .
    \label{eq:7}
\end{equation}

In the same way as the value action function is defined in the Q-learning approach, we define our state action free energy function as follows:

\begin{equation}
    G_t^\pi(x_t, a_t) = \hat{R}_t(x_t, a_t) + \gamma \mathbb{E} \left[F_{t+1}^\pi(x_{t+1}) \mid x_t, a_t \right].
    \label{eq:8}
\end{equation}

By substituting \( F_{t+1}^\pi(x_{t+1}) \) with its value from Eq.~\eqref{eq:6}, we obtain:

\[
G_t^\pi(x_t, a_t) = \hat{R}_t(x_t, a_t) + \gamma \mathbb{E}_{t, a} \left[\sum_{t' = t+1}^{T} \gamma^{t' - t - 1} \left(\hat{R}_{t'}(x_{t'}, a_{t'}) - \frac{1}{\beta} g^\pi(x_{t'}, a_{t'}) \right) \right].
\]

Defining our optimal policy as follows,

\begin{equation}
    \pi(a_t|x_t) = \pi_0(a_t|x_t) e^{\beta (G_t^\pi(x_t, a_t) - F_t^\pi(x_t))}.
    \label{eq:9}
\end{equation}

 we obtain the following expression for \(G\):

\begin{equation}
   G_t^\pi \left( x; a \right) = \hat{R}\left( x_t; a_t \right) + \mathbb{E}_{t,a} \left[\frac{\gamma }{\beta} \log \left( \sum_{a_{t+1}} \pi_0 \left( a_{t+1} | x_{t+1} \right) e^{\beta G_{t+1}^\pi \left( x_{t+1}; a_{t+1} \right)} \right) \right]. 
   \label{eq:10}
\end{equation}

At this level, we need to determine the expression of the reward function  \(\hat{R}_t(x_t, a_t)\) for our example of portfolio optimization for the purchase of a car on a fixed date. So that we can use \(G\) in the sequel to determine our optimal policy.

\subsection*{b. Construction of the reward function}

We consider a discrete-time framework comprising \( T \) steps, where \( T \) represents the time horizon as an integer. The investor or planner manages wealth across \( N \) assets, with \( x_t \) indicating the vector of dollar values assigned to various assets at time \( t \), and \( u_t \) representing the vector of alterations in these positions.

Additionally, we assume that the first asset, identified as \( n = 1 \), is a risk-free bond, while the remaining assets carry risk with uncertain returns \( r_t \), whose expected values are denoted as \( \bar{r}_t \). The returns' covariance matrix is \( \Sigma_r \), which has dimensions of \((N - 1) \times (N - 1)\).

Let \( c_t \) represent the cash contribution made to the plan at time \( t \). Consequently, the combination \( (c_t, u_t) \) can be viewed as the action variables in a dynamic optimization problem related to our problem.

Furthermore, we assume that at each time step \( t \), a predetermined target value \( \hat{P}_{t+1} \) exists for the portfolio at the subsequent time \( t + 1 \).

Let us emphasize again what the optimization task consists of in our problem.
\begin{itemize}
    \item Minimize the contributions required to achieve a certain objective on a given date, while ensuring that the portfolio generates sufficient returns.
    \item Maximize the future value of the portfolio by optimally allocating contributions among different assets, taking into account transaction costs and the risks associated with these choices.
\end{itemize}
To achieve this objective, our reward function will aim to penalize situations where the agent injects capital \( c_t \) into the portfolio. Additionally, it will set a very high target value \( \hat{P}_{t+1} \) and penalize situations where this target value \( \hat{P}_{t+1} \) at step \( t \) exceeds the value at the next step \( V_{t+1} = (1 + r_t)(x_t + u_t) \) of the portfolio (penalty for underperformance relative to this objective).

More formally, we have:
\begin{enumerate}
    \item \textbf{ \(-c_t\): The cost of the contribution. }
    
    This term represents the cost of the cash contribution you make at the beginning of the period. The more you contribute, the higher this cost, which reduces the reward.However, this contribution is necessary to accumulate the capital needed to achieve the objective.

    \item \textbf{ \(-\lambda \mathbb{E}_t \left[(\hat{P}_{t+1} - (1 + r_t)(x_t + u_t))_+\right]\): Penalty for underperformance. }
    
    This term penalizes the gap between the actual performance of the portfolio and the target value set for the next period. If the actual value of the portfolio at the end of the period \( V_{t+1} = (1 + r_t)(x_t + u_t) \) is lower than the target \( \hat{P}_{t+1} \), you receive a penalty proportional to this gap.

      \item \textbf{ \(-u_t^T \Omega u_t\): Transaction costs.}

   This term corresponds to the cost of transactions. Each change in the positions in the portfolio \( u_t \) (buying or selling assets) incurs costs, such as brokerage fees or costs of adjusting the portfolio. \(\Omega\) is a weighting matrix that controls the magnitude of the costs according to the size and type of transactions.
\end{enumerate}

This gives us the following reward function.
\begin{equation}
    R_t \left( x_t, u_t, c_t \right) = -c_t -\lambda \mathbb{E}_t \left[(\hat{P}_{t+1} - (1 + r_t)(x_t + u_t))_+\right] - u_t^T \Omega u_t.
    \label{eq:11}
\end{equation}

However, this function lacks analytical significance because of the rectified non-linearity \( (\cdot)_+ := \max(\cdot, 0) \) within the expectation. We also have the following relation between the variation vector of these positions \( u_t \) and the cash contribution \( c_t \) at time \( t \):

\[
\sum_{n=1}^N u_{tn} = c_t;
\]
So we modify the reward in two ways and we get: 
\begin{equation}
    R_t \left( x_t, u_t \right) = - \sum_{n=1}^N u_{tn} -\lambda \mathbb{E}_t \left[(\hat{P}_{t+1} - (1 + r_t)(x_t + u_t))^2\right] - u_t^T \Omega u_t.
    \label{eq:12}
\end{equation}

The squaring of the expression \(\left( \hat{P}_{t+1} - (1 + r_t)(x_t + u_t) \right)^2\) introduces a symmetry that penalizes both cases where \(V_{t+1} \gg \hat{P}_{t+1}\) and those where \(V_{t+1} \ll \hat{P}_{t+1}\). In reality, our goal is to penalize only the latter scenarios. To address this concern, we can choose target values \(\hat{P}_{t+1}\) that are substantially higher than the expected portfolio value at time \(t\) for the upcoming period. Consequently, we construct this target following the methodology suggested by \cite{Reward}, wherein they describe \(\hat{P}_{t+1}\) as a linear combination of a benchmark \(B_t\) that is independent of the portfolio and a proportional increase in the current portfolio \(\eta 1^T x_t\). Here, \(\eta\) represents the desired growth rate, and \(1^T x_t\) signifies the present portfolio value. Therefore, we obtain:

\begin{equation}
    \hat{P}_{t+1} = (1 - \rho) B_t + \rho \eta 1^T x_t ,
    \label{eq:13}
\end{equation}

Where \(\rho\) is a parameter between 0 and 1, which controls the relative importance of the benchmark component (portfolio-independent) and the portfolio-dependent component.

By representing the asset returns as \(r_t = \bar{r}_t + \tilde{\epsilon}_t\), where the initial component \(\bar{r}_0(t) = r_f\) denotes the risk-free rate (given that the first asset is risk-free), and \(\tilde{\epsilon}_t = (0, \epsilon_t)\), with \(\epsilon_t\) being an idiosyncratic noise, we can substitute Eq.~\eqref{eq:13} into Eq.~\eqref{eq:12} and expand it to derive the following expression:

\begin{equation*}
   \begin{aligned}
       R_t(x_t, u_t) &= -\lambda \hat{P}^2_{t+1} - u_t^T 1 + 2\lambda \hat{P}_{t+1}(x_t + u_t)^T (1 + \bar{r}_t) - \lambda (x_t + u_t)^T \hat{\Sigma}_t (x_t + u_t) - u_t^T \Omega u_t\\
      &= x_t^T R^{(xx)}_t x_t + u_t^T R^{(ux)}_t x_t + u_t^T R^{(uu)}_t u_t + x_t^T R^{(x)}_t + u_t^T R^{(u)}_t + R^{(0)}_t .
   \end{aligned} 
\end{equation*}
  
where
\begin{equation*}
    \begin{aligned}
        \hat{\Sigma}_t &= \begin{pmatrix} 0 & 0 \\ 0 & \Sigma_r \end{pmatrix} + (1 + \bar{r}_t)(1 + \bar{r}_t)^T \\
         R^{(xx)}_t &= -\lambda \eta^2 \rho^2 11^T + 2\lambda \eta \rho (1 + \bar{r}_t)1^T - \lambda \hat{\Sigma}_t \\
        R^{(ux)}_t &= 2\lambda \eta \rho (1 + \bar{r}_t)1^T - 2\lambda \hat{\Sigma}_t \\
        R^{(uu)}_t &= -\lambda \hat{\Sigma}_t - \Omega \\
        R^{(x)}_t &= -2\lambda \eta \rho (1 - \rho) B_t 1 + 2\lambda (1 - \rho) B_t (1 + \bar{r}_t) \\
        R^{(u)}_t &= -1 + 2\lambda (1 - \rho) B_t (1 + \bar{r}_t) \\
        R^{(0)}_t &= -(1 - \rho)^2 \lambda B_t^2 .
    \end{aligned}
\end{equation*}

If we consider the expected returns \(\bar{r}_t\), the covariance matrix \(\Sigma_r\), and the benchmark \(B_t\) as constant, the vector of free parameters that characterize the reward function can be expressed as \(\theta := (\lambda, \eta, \rho, \Omega)\).

At this stage, our objective is to determine the optimal policy. The G-learner algorithm, whose mathematical approach is described below, uses the previously defined functions \(F\) and \(G\) to achieve this goal.

\subsection*{c. Optimal Policy Determination: G-learner}

G-learner is an algorithm proposed by \cite{dixon2020machine} for optimal policy determination in G-learning. In this section, we describe the steps of this algorithm.

In order to facilitate the exploitation of the free energy function \(F^\pi_t\), it is necessary to transform it into a quadratic functional form. This transformation is made possible by the quadratic nature of the reward function. Thus, we obtain:  
\begin{equation}
    F^\pi_t(x_t) = x_t^T F^{(xx)}_t x_t + x_t^T F^{(x)}_t + F^{(0)}_t ,
    \label{eq:14}
\end{equation}

where \(F^{(xx)}_t\), \(F^{(x)}_t\), and \(F^{(0)}_t\) are parameters that can depend on time via their dependence on the target values \(\hat{P}_{t+1}\) and the expected returns \(\bar{r}_t\).

To determine the coefficients of this expression, we also transform the G function into a quadratic form as follows: 
\[
 G_t^{\pi}(x_t, u_t) = x_t^\top Q_t^{(xx)} x_t + u_t^\top Q_t^{(ux)} x_t + u_t^\top Q_t^{(uu)} u_t + x_t^\top Q_t^{(x)} + u_t^\top Q_t^{(u)} + Q_t^{(0)}.
\] 
Here, \( Q_t^{(xx)}\), \(Q_t^{(ux)}\),\(Q_t^{(uu)}\),\(Q_t^{(x)}\),\(Q_t^{(u)}\), and \(Q_t^{(0)}\) are easily determined by expanding the equation Eq.~\eqref{eq:10}.

We also assume that the reference policy \(\pi_0\) is defined as:  
\begin{equation}
   \pi_0(u_t \mid x_t) = \frac{1}{\sqrt{(2\pi)^n \lvert \Sigma_p \lvert}} e^{-\frac{1}{2} (u_t - \hat{u}_t)^\top \Sigma_p^{-1} (u_t - \hat{u}_t)},
   \label{eq:15}
\end{equation}

where \(\hat{u}_t\) is an adjusted mean as follows:  
\(
\hat{u}_t = \bar{u}_t + \bar{v}_t x_t 
\)
These elements allow us to obtain the following coefficients for the function \(F^\pi_t\).
\begin{equation*}
    \begin{aligned}
       F_t^{(xx)} &= Q_t^{(xx)} + \frac{1}{2\beta} \left(U_t^\top \Sigma_p^{-1} U_t - \bar{v}_t^\top \Sigma_p^{-1} \bar{v}_t\right)\\
       F_t^{(x)} &= Q_t^{(x)} + \frac{1}{\beta} \left(U_t^\top \Sigma_p^{-1} W_t - \bar{v}_t^\top \Sigma_p^{-1} \bar{u}_t\right) \\
       F_t(0) &= Q_t(0) + \frac{1}{2\beta} \left(W_t^\top \Sigma_p^{-1} W_t - \bar{u}_t^\top \Sigma_p^{-1} \bar{u}_t\right) - \frac{1}{2\beta} \log |\Sigma_p| + \log |\bar{\Sigma}_p|.
    \end{aligned}
\end{equation*}
where we use the auxiliary parameters:

\begin{equation*}
    \begin{aligned}
        U_t = \beta Q_t^{(ux)} + \Sigma_p^{-1} \bar{v}_t
        W_t = \beta Q_t^{(u)} + \Sigma_p^{-1} \bar{u}_t
        \bar{\Sigma}_p = \Sigma_p^{-1} - 2\beta Q_t^{(uu)} .
    \end{aligned}
\end{equation*}
The optimal policy for the given step is:
\begin{equation}
    \pi(u_t | x_t) = \pi_0(u_t | x_t) e^{\beta \left(G^\pi_t(x_t, u_t) - F^\pi_t(x_t)\right)}.
    \label{eq:16}
\end{equation}

By substituting \(\pi_0(u_t | x_t)\) and \(G^\pi_t(x_t, u_t)\) into this expression, we obtain:
\[
\pi(u_t | x_t) = \frac{1}{(2\pi)^{n/2} |\tilde{\Sigma}_p|^{1/2}} e^{-\frac{1}{2}(u_t - \tilde{u}_t - \tilde{v}_t x_t)^\top \tilde{\Sigma}_p^{-1}(u_t - \tilde{u}_t - \tilde{v}_t x_t)} .
\]
where
\begin{equation*}
    \begin{aligned}
        \tilde{\Sigma}_p^{-1} &= \Sigma_p^{-1} - 2\beta Q_{uu,t}\\
        \tilde{u}_t &= \tilde{\Sigma}_p(\Sigma_p^{-1} \bar{u}_t + \beta Q_{u,t})\\
        \tilde{v}_t &= \tilde{\Sigma}_p(\Sigma_p^{-1} \bar{v}_t + \beta Q_{ux,t}) .
    \end{aligned}
\end{equation*}

The process necessary for learning \(G\) in this context includes determining the optimal liquidity contribution at each step while adhering to the budget constraint. Given that the learning \(G\) generates Gaussian actions \(u_t\), the optimal contribution at time \(t\), denoted as \(c_t\), is expected to follow a Gaussian distribution with a mean of \(\bar{c}_t = 1^\top (\bar{u}_t + \bar{v}_t x_t)\). Thus, the expected optimal contribution \(\bar{c}_t\) comprises a component \(\sim \bar{u}_t\) that is independent of the portfolio value, alongside a component \(\sim \bar{v}_t\) that is contingent on the current portfolio.

Since the reward function depends on several parameters, two approaches can be considered for applying our G-learner:
\begin{enumerate}
    \item \textbf{Direct Parameter Setting:}
    \begin{itemize}
        \item In this first approach, the parameters of the reward function are directly defined.
        \item G-learner is then applied to find the optimal policy.
    \end{itemize}

    \item \textbf{Parameter Learning via Imitation:}
    \begin{itemize}
        \item The second approach involves learning the parameters by imitating the behavior of a G-learning agent that has produced good results.
        \item The parameters are optimized before applying the G-learner algorithm to determine the optimal policy.
        \item The algorithm used for this approach is known as \textbf{GIRL}.
    \end{itemize}
\end{enumerate}

\subsection*{d. Learning reward function parameters: GIRL}

Application of the G-learner framework is appropriate when the investor clearly specifies his reward function and knows the values of the parameters \(\theta := (\lambda, \eta, \rho, \Omega)\). However, in many scenarios, as is currently the case, the reward function remains unknown and depends on a set of parameters. Such situations, where rewards are not accessible, fall within the domain of inverse reinforcement learning (IRL), which aims to determine both the agent's reward function and the optimal policy.

We seek to demonstrate how to deduce the reward function based on the observed actions of an agent. For this purpose, we assume access to a collection of trajectories. Furthermore, we possess historical data on asset prices and anticipated returns for every asset within the investor's universe.

The principle is therefore to start from this set of supposedly independent trajectories, generated by a G-learner agent in our case, and try to reconstitute the hidden reward function by minimising the log likelihood of these trajectories. Mathematically, we have:

Assume we possess a collection of \(D\) trajectories \(\zeta_i\) where \(i = 1, \ldots, D\), each consisting of state-action pairs \((x_t, u_t)\). Each trajectory \(i\) begins at a specific time \(t_0^i\) and continues until time \(T_i\). Now, consider one trajectory \(\zeta\) from this set, setting its starting point at \(t = 0\) and its endpoint at \(T\). We posit that the dynamics are Markovian with respect to the pair \((x_t, u_t)\), described by a generative model given by 
\(p_\theta(x_{t+1}, u_t \mid x_t) = \pi_\theta(u_t \mid x_t) p_\theta(x_{t+1} \mid x_t, u_t)\),  where \(\Theta\) represents a vector of model parameters, and \(\pi_\theta\) is the action policy defined by Eq.~\eqref{eq:9}.

The likelihood of witnessing trajectory \(\zeta\) can be expressed as follows:

\begin{equation}
    P \left(x, u \mid \Theta\right) = p_0\left(x_0\right) \prod_{t=0}^{T-1} \pi_\theta\left(u_t \mid x_t\right) p_\theta \left(x_{t+1} \mid x_t, u_t\right),
    \label{eq:17}
\end{equation}

\begin{equation}
    {LL}\left(\theta\right) := \log P \left(x, u \mid \Theta\right) = \sum_{t \in \zeta} \left(\log \pi_\theta\left(u_t \mid x_t\right) + \log p_\theta \left(x_{t+1} \mid x_t, u_t\right)\right).
    \label{eq:18}
\end{equation}

Using the expression \(x_{t+1} = A_t (x_t + u_t) + (x_t + u_t) \circ \tilde{\epsilon}_t\) with \(A_t := \text{diag}(1 + \bar{r}_t)\), we can express the transition probability as follows:

\begin{equation}
    p_\theta \left(x_{t+1} \mid x_t; u_t\right) = \frac{e^{-\frac{1}{2} \Delta_{t}^T \Sigma_r^{-1} \Delta_t} }{\sqrt{\left(2\pi\right)^{N} |\Sigma_r|}} \delta \left(x^{(0)}_{t+1} - \left(1 + r_f\right)x^{(0)}_t\right),
    \label{eq:19}
\end{equation}

\[
\Delta_t := \frac{x^{(r)}_{t+1} }{x^{(r)}_t + u^{(r)} _t} -{A^{(r)}_t}.
\]

The term \(\delta \left(x^{(0)}_{t+1} - \left(1 + r_f\right)x^{(0)}_t\right)\) represents the deterministic behavior of the bond component within the portfolio. 

These expressions allow us to derive the log-likelihood that the trajectory \( \zeta \) was generated by the parameters \( \pi_\theta \) as follows:
 
\[
LL\left(\theta\right) = \sum_{t \in \zeta} \left[ \beta G_t^\pi\left(x_t; u_t\right) - F_t^\pi\left(x_t\right) - \frac{1}{2} \log |\Sigma_r| - \frac{1}{2} \Delta_t^T \Sigma_r^{-1} \Delta_t \right], 
\]

Similarly, the likelihood of all the trajectories in the set \(D\) is computed. The loss function to be used will then be obtained by summing these log-likelihoods.

\subsubsection{Implementation Process of G-learning }
To implement our G-learning, we adopt two main approaches. The first approach involves defining the parameters of the reward function, using a G-learner agent to determine the optimal policy, and finally constructing the optimal portfolio. The second approach is based on inverse reinforcement learning, where the parameters of the reward function are learned using the GIRL algorithm, followed by the same steps as in the first approach. In these approaches, the key algorithms used are G-learner and GIRL. Below, we outline the different steps of the G-learning implementation before describing these two algorithms.
The following steps outline the implementation of our G-learning model:

\begin{itemize}
    \item Initialize the parameters of the initial policy (prior).
    \item Initialize the reference portfolio (\texttt{benchmark\_portf}).
    \item Invoke the G-learner to determine the optimal policy if the parameters are already known; otherwise, use GIRL first to learn the parameters.
    \item Generate the action \(u_t\) by sampling from a normal distribution with mean \(\mu_t\) and covariance \(\Sigma_{\text{prior}}[t]\).
    \item Update the portfolio state.
    \item Simulate the returns of risky assets based on a one-factor model.
    \item Update the returns and states.
    \item Add \((x_t, u_t)\) to the current trajectory.
    \item Store the trajectory and returns in the respective lists.
\end{itemize}

\section*{a.G-Learner Algorithm}

The G-Learner is an algorithm that computes an optimal policy using the \(G\) and \(F\) functions. Below are its main steps:

\begin{itemize}
    \item \textbf{Initialization:} Both \(G\) and \(F\) functions are initially set to zero.
    \item \textbf{Calculate the reward function:}  for each given time t.
    \item \textbf{Updating \(G\) and \(F\) Values:} These values are updated at every step, aiming to maximize long-term rewards.
    \item Update the policy at each time step .
    \item \textbf{Convergence or Iteration Limit:} The algorithm stops either when the \(G\) and \(F\) functions converge or when a maximum number of iterations is reached.
\end{itemize}
Algorithm for G-learner implementation.

\begin{algorithm}[H]
\caption{G-learner Algorithm} \label{alg:G-leaner}
 \begin{algorithmic}[1]
  \STATE \textbf{Input}:
    \begin{itemize}
    \item \texttt{num\_steps}: number of time steps
    \item \texttt{x\_vals\_init}: initial state (initial portfolio value)
    \item \texttt{reward\_params}: parameters of the reward function
    \item \texttt{beta}: discount factor
    \item \texttt{gamma}: risk aversion
    \item \texttt{num\_risky\_assets}: number of risky assets
    \item \texttt{riskfree\_rate}: risk-free rate
    \item \texttt{expected\_risky\_returns}: expected returns of risky assets
    \item \texttt{Sigma\_r}: covariance matrix of risky assets
    \item \texttt{max\_iter\_RL}: maximum number of iterations for convergence
    \item \texttt{eps}: convergence tolerance
  \end{itemize}
  \STATE \textbf{Output}: Optimal policy \( \pi^*(x_t) \) and the optimized functions \( G^* \) and \( F^* \).
 \STATE Initialize the values for \( G_t(x_t, u_t) \) and \( F_t(x_t) \) to 0 for all \( t \), \( x_t \), and \( u_t \).\;
  \FOR{ $n$ from $0$ to \texttt{max\_iter\_RL}}
         \FOR{$t$ from $0$ \texttt{num\_steps}}
             \STATE Calculate the reward function:
                    \begin{equation*}
                        \begin{aligned}
       R_t(x_t, u_t) &= -\lambda \hat{P}^2_{t+1} - u_t^T 1 + 2\lambda \hat{P}_{t+1}(x_t + u_t)^T (1 + \bar{r}_t) - \lambda (x_t + u_t)^T \hat{\Sigma}_t (x_t + u_t) - u_t^T \Omega u_t\\
      &= x_t^T R^{(xx)}_t x_t + u_t^T R^{(ux)}_t x_t + u_t^T R^{(uu)}_t u_t + x_t^T R^{(x)}_t + u_t^T R^{(u)}_t + R^{(0)}_t .
                    \end{aligned} 
              \end{equation*}
       \STATE Update the functions \(G\) and \(F\) using the backward Bellman algorithm to optimize future actions:
                          \begin{itemize}
                                
                                \item \(  G_t^{\pi}(x_t, u_t) = x_t^\top Q_t^{(xx)} x_t + u_t^\top Q_t^{(ux)} x_t + u_t^\top Q_t^{(uu)} u_t + x_t^\top Q_t^{(x)} + u_t^\top Q_t^{(u)} + Q_t^{(0)} \).
                                \item \(  F_t(x_t) = \max_{u} G_t(x_t, u) \).
                           \end{itemize}

         \STATE Update the policy at each time step:
          \[
          \pi(u_t | x_t) = \frac{1}{(2\pi)^{n/2} |\tilde{\Sigma}_p|^{1/2}} e^{-\frac{1}{2}(u_t - \tilde{u}_t - \tilde{v}_t x_t)^\top \tilde{\Sigma}_p^{-1}(u_t - \tilde{u}_t - \tilde{v}_t x_t)} .
        \]          
        
     \ENDFOR
     \STATE Exit the iteration loops once the tolerated error (\( \text{error\_tol} \)) is reached
   \ENDFOR  
    \end{algorithmic}
\end{algorithm}

\section*{b.GIRL Algorithm}

The GIRL (Gradient Inverse Reinforcement Learning) algorithm infers the reward function based on observed behaviors. It aims to minimize a loss function that measures the discrepancy between observed trajectories and those generated by the estimated reward function. Below are the main steps of its implementation:

\begin{itemize}
    \item \textbf{Initialization of Parameters:}  
    Initialization of the reward function parameters \( \theta \).
    \item \textbf{Policy Estimation:}  
    GIRL utilizes the G-Learner algorithm to compute the optimal policy \( \pi_{\theta} \) for the current reward parameters.

    \item \textbf{Likelihood Calculation:}  
    For each observed trajectory, GIRL evaluates the probability that this trajectory was generated by the policy \( \pi_{\theta} \).

    \item \textbf{Loss Minimization:}  
    GIRL updates \( \theta \) by minimizing the loss function (negative likelihood of the observed trajectories) using the gradient descent method.
    \item \textbf{Convergence:} The process is repeated until the policy \( \pi_{\theta} \) is sufficiently close to the one that generated the observed trajectories.

\end{itemize}

Algorithm for GIRL implementation.
\begin{algorithm}[H]
\caption{Inverse Reinforcement Learning (GIRL) Algorithm}\label{alg:GIRL}
\begin{algorithmic}[1]
    \STATE \textbf{Input:} 
    \begin{itemize}
        \item \texttt{\( \tau \) }: observed trajectories (observed states and actions)
        \item \texttt{num\_sim}: number of simulation
        \item \texttt{num\_steps}: number of time steps
        \item \texttt{initial\_reward\_params}: initial parameters of the reward function
        \item \texttt{beta}: discount factor
        \item \texttt{gamma}: risk aversion
        \item \texttt{max\_iter\_RL}: maximum number of iterations for G-Learner
        \item \texttt{eps}: tolerance for gradient (for gradient calculations)
        \item \texttt{learning\_rate}: learning rate for updating reward function parameters
    \end{itemize}
    \STATE \textbf{Output:} Optimal reward function parameters \( \theta^* \) explaining observed behaviors.
    \STATE Initialize \( \theta = \text{initial\_reward\_params} \) (initial reward function parameters).
    \FOR{each iteration until convergence or reaching the maximum number of iterations}
        \STATE Use the G-Learner to compute the policies \( \pi_{\theta} \) for the parameters \( \theta \).
        \FOR{each observed trajectory \( \tau = \{(x_t, u_t)\}_{t=1}^{T} \)}
            \STATE Compute log-likelihood  \( LL\left(\theta\right) = \sum_{t \in \zeta} \left[ \beta G_t^\pi\left(x_t; u_t\right) - F_t^\pi\left(x_t\right) - \frac{1}{2} \log |\Sigma_r| - \frac{1}{2} \Delta_t^T \Sigma_r^{-1} \Delta_t \right] \) that \( \pi_{\theta} \) generated this trajectory.
        \ENDFOR
        \STATE Compute the loss function (sum negative log-likelihood):
                 \[{Loss}(\theta) = - \sum_{\zeta \in \tau}\sum_{t \in \zeta} \left[ \beta G_t^\pi\left(x_t; u_t\right) - F_t^\pi\left(x_t\right) - \frac{1}{2} \log |\Sigma_r| - \frac{1}{2} \Delta_t^T \Sigma_r^{-1} \Delta_t \right]\] 
           \STATE Update the reward function parameters:
             \begin{itemize}
                 \item Compute the gradients $\nabla_{\theta} {Loss}(\theta)$ of the loss function with respect to \( \theta \).
                 \item  \[
            \theta \leftarrow \theta - \text{learning\_rate} \cdot \nabla_{\theta} {Loss}(\theta)
            \]
             \end{itemize}
        \STATE Check for convergence (if the loss function or gradients are close to a tolerance \( \epsilon \)).
    \ENDFOR
\end{algorithmic}
\end{algorithm}

\subsection{Description of data and simulation principle}

Our study aims to demonstrate the effectiveness of reinforcement learning in solving real-world portfolio optimization problems with defined contributions. To achieve this, we rely on popular financial models, known for their ability to accurately replicate the dynamics of financial markets, to generate our data.

\subsubsection{Market and Asset Modeling}
To model our market, we start with the assumption that the evolution of assets follows a Markov process. This means that the future price of an asset depends solely on its current price, without considering its historical price. This assumption accurately reflects the reality of financial markets. In fact, it is characteristic of geometric Brownian motion (GBM), which is widely used in mathematical finance to model stock prices, particularly in the Black-Scholes model. This model is the most commonly used to represent the behavior of stock prices and, more generally, that of financial markets.

Mathematically, this process is defined by the following relationship:
     \[
S_t = S_{t-1} \times \exp\left(\left(\mu - \frac{1}{2} \sigma^2\right) \Delta t + \sigma \sqrt{\Delta_t} \, Z\right)
\]

Where:
\begin{itemize}
    \item \( S_t \) is the market value at time \( t \).
    \item \( S_{t-1} \) is the market value at the previous time step \( t-1 \).
    \item \( \mu \) is the drift term or the expected return of the market.
    \item \( \sigma \) is the volatility or standard deviation of the market returns.
    \item \( \Delta_t \) is the time increment.
    \item \( Z \) is a random variable drawn from a standard normal distribution, i.e., \( Z \sim \mathcal{N}(0,1) \).
\end{itemize}

For the market return at time t, denoted $r_t$, we use a log-return to model it. This gives the following relationship.
\[
\quad r_t = \nu_M \Delta_t + \sigma \sqrt{\Delta t} Z_t
\]
Where : $\nu_M$ is market drift

For risky assets, we have two developments of returns over time: one is the expected return, and the other is the return generated due to market impact (realized return)

\subsection*{a.The expected risky asset returns, \(\bar{r}_{t,i}\), are given by:}

\[
r_{t,i} = \bar{r}_{t,i} + \beta^0_i (r_M - \mu_M dt) + \sigma_i \sqrt{1 - (\beta^0_i)^2} dW_{t,i}, \quad i \in \{1, \dots, N - 1\},
\]
where \(\mu_M = 0.05\) represents the drift of the market, \(r_M\) denotes the market returns generated under a geometric Brownian motion (GBM) framework with a volatility of \(\sigma_M = 0.25\). Additionally, \(\beta^0_i\) signifies the beta coefficient of the \(i\)-th asset. The term \(\sigma_i \equiv \sigma = 0.05\) indicates the idiosyncratic volatility, and \(dW_t\) is a Brownian motion that drives the system and is correlated with the market noise, with \(dt = 0.25\).

\subsection*{b.realized returns \(\bar{r}_t\) is assumed to be given by the CAPM:}

\[
\bar{r}_t = \alpha + \beta^0 \left[(1 - c)\mu_M dt + cr_M \right], \quad c \in [0, 1],
\]
where we select the oracle coefficient \(c = 0.2\).

We consider \(\alpha\) and \(\beta^0\) as uniformly distributed random variables across all risky assets, defined as:
\[
\alpha \sim U([-0.05, 0.15]), \quad \beta^0 \sim U([0.05, 0.85]).
\]

These relationships enable the simulation of returns and values for risky assets within a single-factor model, where the returns are influenced by both market returns (weighted by a beta coefficient unique to each asset) and an idiosyncratic component (the specific volatility associated with each asset).

\subsubsection{Simulation}
We therefore use the information from Tab.~\ref{table:finance_variables} below to simulate our data. This information is mainly related to the initial value of our market, the number of risky assets, the duration of our investment, the market return, the market volatility, and the risk-free rate.

\begin{table}[H]
\centering
\begin{tabular}{|>{\centering\arraybackslash}p{7cm}|>{\centering\arraybackslash}p{5cm}|}
\hline
\textbf{Financial Context Name} & \textbf{Value} \\ \hline
\texttt{Expected Market Return} & $0.05$ \\ \hline
\texttt{Market Volatility} & $0.25$ \\ \hline
\texttt{Initial Market Value} & $100.0$ \\ \hline
\texttt{Risk-Free Rate} & $0.02$ \\ \hline         
\texttt{Number of Time Steps} & $25$ \\ \hline
\texttt{Time Increment (Quarterly)} & $0.25$ \\ \hline
\texttt{Number of Risky Assets} & $99$ \\ \hline
\end{tabular}
\caption{Table of market and asset modeling parameters.}
\label{table:finance_variables}
\end{table}
We can therefore observe the behavior of our market over time (based on variations in the Return of this market and variations in its value).

\begin{figure}[H]
    \centering
    \includegraphics[width=0.8\linewidth]{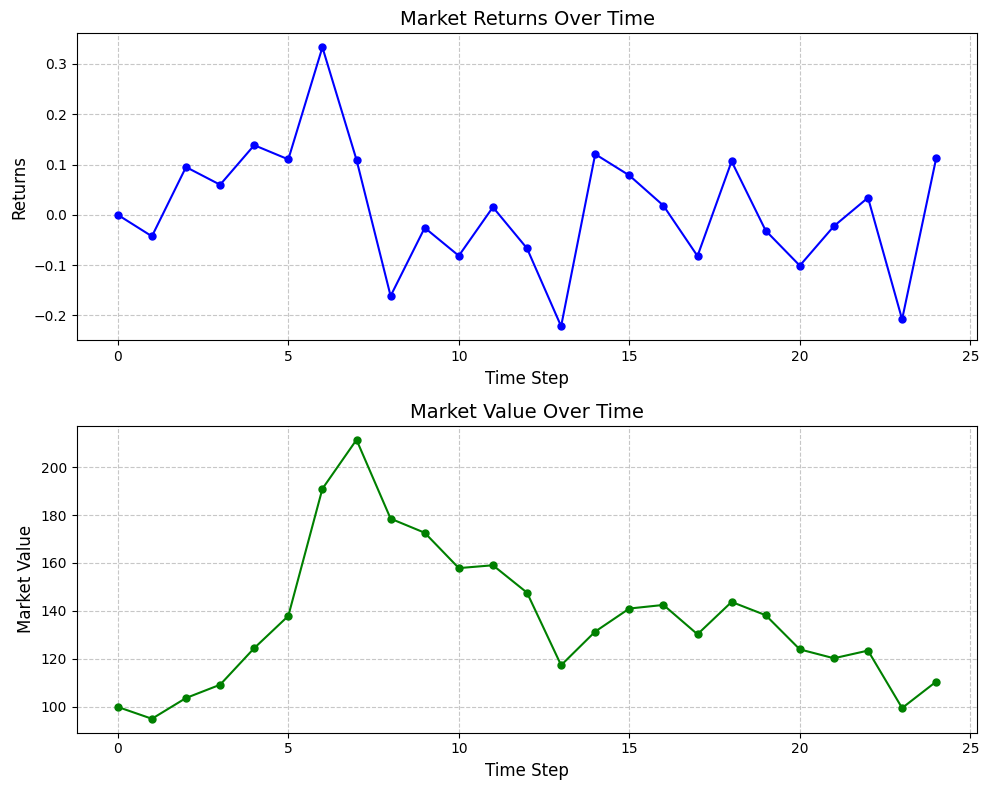}
    \caption{Variation of Market returns and Market value over time }
    \label{fig:Market}
\end{figure}

We can also observe the behavior of some simulated risky assets on this market.
\begin{figure}[H]
    \centering
    \includegraphics[width=1\linewidth]{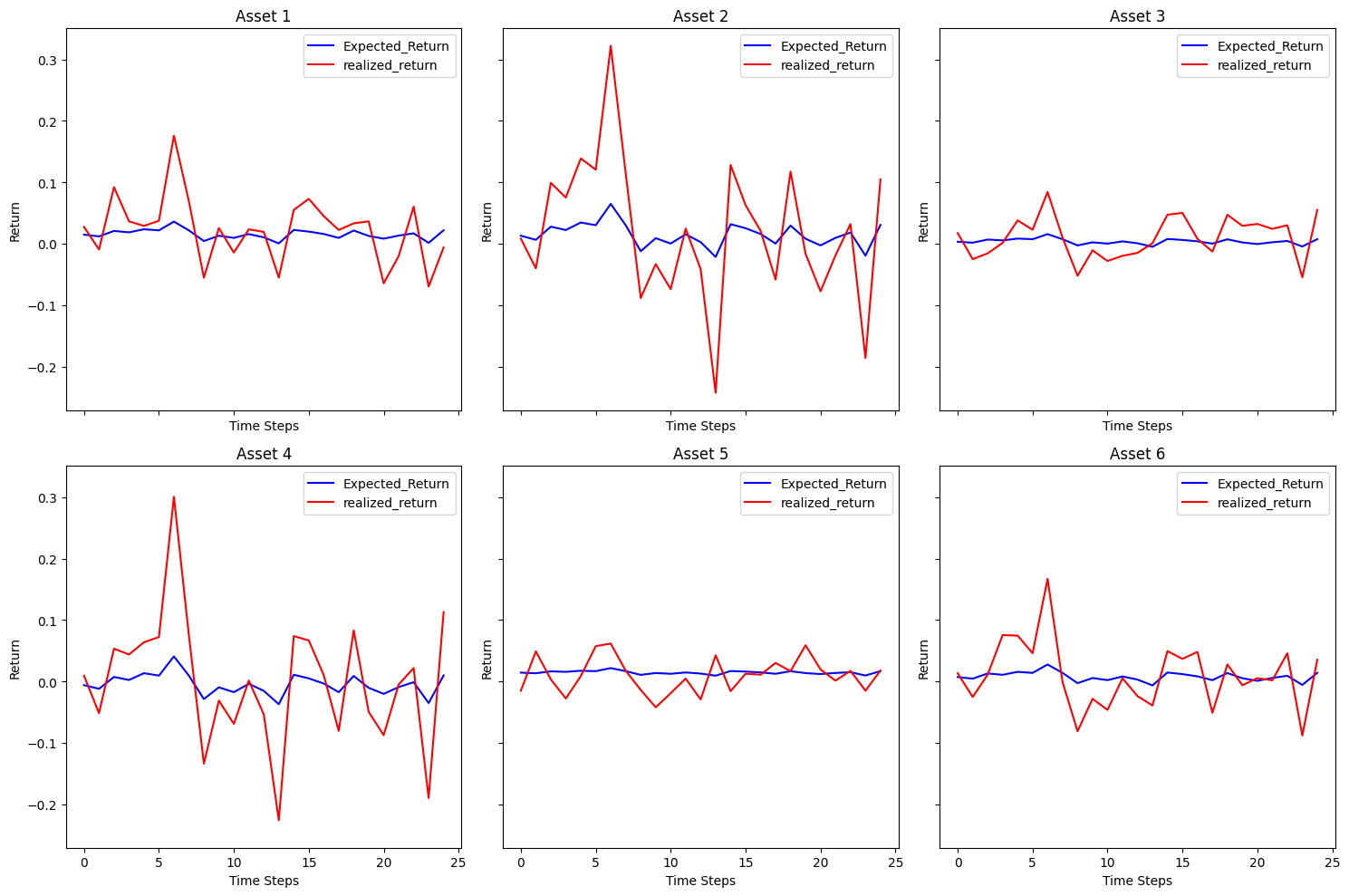}
    \caption{This figure shows the difference between the realized return and the expected return on a number of risky assets.}
    \label{fig:assets}
\end{figure}

Fig.~\ref{fig:Market} shows a significant variation in the return and value of our market over time, which is generally the case in most financial markets. We can directly observe the impact of this high market volatility on the realized returns of our assets, which completely deviate from the expected returns, as shown in Fig.~\ref{fig:assets}.

The main task of our reinforcement learning model is to learn how to anticipate sudden changes in the returns of various assets caused by market volatility. The goal is to adjust the portfolio before each new fluctuation to maintain the balance required to reach the target at the defined horizon.

The results obtained with our model will be presented in the next section.

\subsection{Results} 

\subsection{Presentation of G-learning and GIRL performances } 
\subsubsection{Portfolio performance after direct optimization with G-learning}
After applying a set of random parameters in a well-defined range to our G-learner , we obtained the best result with the following parameter values: \(\theta := (\lambda = 0.002, \eta = 1.3, \rho = 0.5, \Omega = 1.1)\)

\begin{figure}[H]
    \centering
    \includegraphics[width=0.5\linewidth]{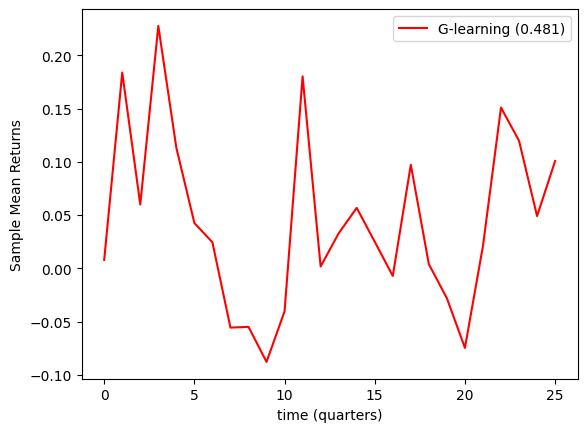}
    \caption{Adjusted portfolio returns over time using the G-learning model with fixed parameters.}
    \label{fig:G-learning}
\end{figure}

Indeed, with these parameters, our G-learning model was able to adjust the evolution of our portfolio's return to produce a Shape Ratio of $= 0.481$. According to the work of \cite{schmid2010statistical}, in high-volatility financial markets, Shape Ratio values vary between $0.2$ and $0.7$, and a Shape Ratio value of $0.3$ is already considered relevant. In view of the high volatility of our market, we can conclude that our model produces a fairly interesting result.  Fig.~\ref{fig:G-learning} shows the evolution of the portfolio return optimized by our G-learner money and the value of the shape ratio obtained.

\subsubsection{Parameters learned by the GIRL algorithm}

We then use the G-learner agent, developed during the previous G-learning implementation, to simulate a set of trajectories (representing the changes in portfolio positions over time). The GIRL algorithm leverages these trajectories to infer and optimize the parameters of the reward function that was used to build the G-learner agent. Fig.~\ref{fig:loss} below shows the values of the optimal parameters obtained. Tab.~\ref{tab:comparer} compares the values learned by GIRL with the initial parameters used to construct our G-learner agent, highlighting GIRL's ability to reconstruct and refine a reward function.

\begin{table}[H]
\centering
\caption{Comparison of Parameters Learned by  GIRL}
\label{tab:comparer}
\begin{tabular}{|>{\centering\arraybackslash}p{3.5cm}|>{\centering\arraybackslash}p{3.5cm}|>{\centering\arraybackslash}p{3.5cm}|}
\hline
  \textbf{Parameters}& \textbf{G-learning} & \textbf{ GIRL} \\ \hline
   $\lambda$ & 0.002 & 0.0012 \\ \hline
   $\omega$ & 1.1 & 1.01 \\ \hline
    $\eta$ & 1.3 & 1.5 \\ \hline        
    $\rho$ & 0.5 & 0.4 \\ \hline
  \end{tabular}
\end{table} 

\begin{figure}[H]
    \centering
    \includegraphics[width=0.75\linewidth]{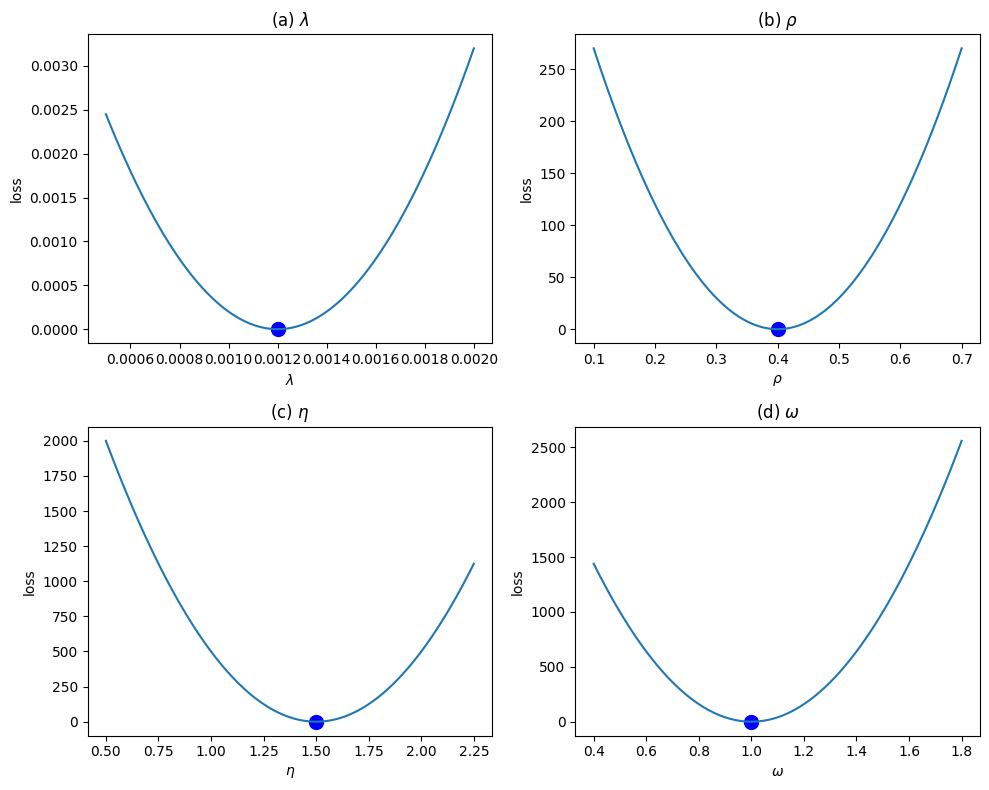}
    \caption{Parameters learning curves using the GIRL model.}
    \label{fig:loss}
\end{figure}

\subsubsection{Comparison of G-learning and GIRL returns}
The Fig.~\ref{fig:GIRL and G-learning} below shows that the difference in performance between the portfolio optimized by our initial G-learning model and the one optimized using the parameters estimated by GIRL is negligible. However, we observe that GIRL achieves a slightly higher Sharpe ratio.

\begin{figure}[H]
    \centering
    \includegraphics[width=0.6\linewidth]{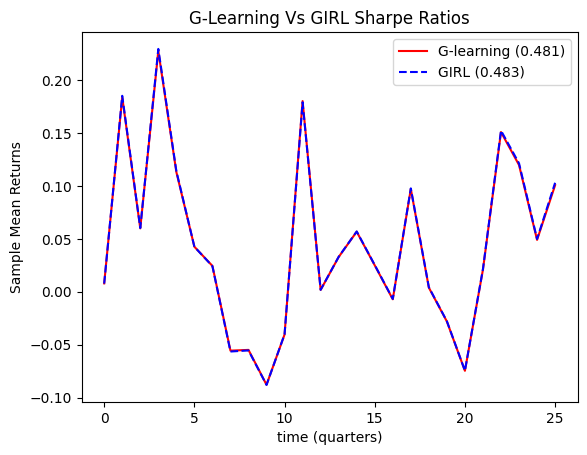}
    \caption{Adjust portfolio returns over time using the G-learner model.}
    \label{fig:GIRL and G-learning}
\end{figure}

\subsection{Presentation of portfolio optimisation results}
For the analytical aspects of our reward function, recall that the reference portfolio (benchmark) was designed to follow exponential growth. As the value of our portfolio increases, the objective becomes progressively more ambitious. This approach also offers an advantage: it allows our portfolio to grow rapidly, even if the agent struggles to keep pace with the benchmark's performance over time. Fig. ~\ref{fig:portfolio} below illustrates the evolution of our portfolio over time.

\begin{figure}[H]
    \centering
    \includegraphics[width=0.8\linewidth]{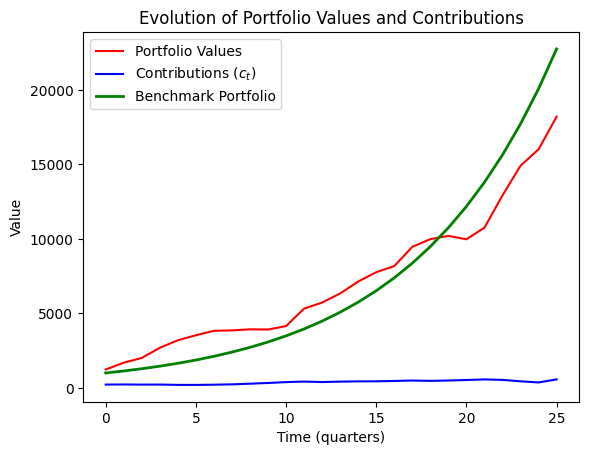}
    \caption{Evolution of the portfolio over time and regular contributions $C_t$}
    \label{fig:portfolio}
\end{figure}
Applying our model to a specific case of a portfolio with an initial value of \$1000, equally distributed across our various assets at time \( t = 0 \), reveals a significant growth in portfolio value, closely aligning with the investor’s target (benchmark portfolio) at the end of the investment period. This was achieved with minimal investor contributions \( c_t \) at each time step \( t \). We can thus conclude that our model effectively establishes a robust portfolio management strategy, allowing us to efficiently manage assets over time to reach our objectives namely, maximizing portfolio value while minimizing investor contributions \( c_t \) throughout the period.

In the following sections, we will analyze our approach and the results obtained, describe future research directions, and provide a general conclusion for the overall work conducted.

\section{Discussion}
The results of this research show that portfolio optimization using G-Learning and the GIRL algorithm effectively maximizes portfolio value at the target date while minimizing regular contributions. The developed model achieved a Sharpe ratio of 0.483, which, according to \cite{schmid2010statistical}, is significant for a highly diversified portfolio in markets characterized by high volatility. This ratio indicates that our method provides a good balance between return and risk in an uncertain environment, producing a relatively high Sharpe ratio with a well-diversified portfolio and, therefore, a reduced risk level.

The success of G-Learning lies in its ability to account not only for market fluctuations but also for periodic contributions. Compared to traditional approaches based on supervised learning or classical techniques such as Modern Portfolio Theory, our method is more flexible and dynamic. Our results also align with the findings of \cite{dixon2020g}, demonstrating that reinforcement learning-based techniques are better suited to complex markets and highly diversified portfolios.

However, we found that the optimal parameters obtained using the GIRL model (e.g., $\lambda = 0.0012$ compared to an initial value of 0.002) result in only a marginal improvement in the Sharpe ratio. This suggests that although GIRL effectively adjusts the reward function parameters, its overall impact on portfolio performance remains limited. A possible reason for this performance is that the existing agent is already efficient and therefore simulates portfolio evolutions that are quite similar, reducing the effect of parameter adjustments on GIRL's overall profitability. It tends to produce portfolio evolutions close to those simulated, leading to a minimal optimization of the Sharpe Ratio. It would be interesting to test this approach in less volatile markets or with longer investment horizons to better evaluate the impact of the optimized parameters.

Our results also emphasize that the goal-based formulation (focused on specific objectives) is better aligned with investors' needs than Markowitz's classical approaches. By considering specific financial objectives at a given date, our model provides more targeted management, in line with the work of \cite{browne2000stochastic} and \cite{das2020dynamic}.

\section{Conclusion}

This research proposes an enhancement of the innovative portfolio optimization approach based on the G-Learning algorithm and parametric optimization using GIRL, as introduced by \cite{dixon2020machine}. The objective was to maximize portfolio value by a target date while minimizing periodic contributions from the investor within a highly diversified portfolio. The results demonstrate that our method achieves a significant Sharpe ratio in a volatile market environment, underscoring the relevance of reinforcement learning for dynamic financial problems. Our study offers a solution to goal-based optimization, enabling investment decisions to align more closely with the investor’s specific needs, such as funding a purchase by a given date. This approach surpasses the limitations of traditional models like Modern Portfolio Theory by incorporating greater flexibility and adaptability to market shifts. The importance of this research lies in demonstrating that probabilistic reinforcement learning is not only applicable but also effective in asset management. By providing a framework that considers both regular contributions and market fluctuations, this method serves as a valuable tool for portfolio managers and individual investors .

\subsection{Recommendations}

For future research, we first suggest exploring the construction of a reward function that could further enhance performance, or alternatively, considering a different approach to regularizing Q-learning. Secondly, we recommend applying G-Learning to other asset classes and in various macroeconomic contexts. Integrating exogenous data and developing real-time adaptive approaches could also open promising new avenues.





\section{Acknowledgments}

This work constitutes my final thesis at the African Institute of Mathematical Sciences (AIMS) in South Africa and Stellenbosch University. I would first like to express my deep gratitude to my supervisors, \textbf{Dr. Rock Stephane Koffi} and \textbf{Dr. Prudence Djagba}, for the confidence they placed in me, as well as for their support, guidance, comments, and encouragement throughout the writing of this thesis. I am deeply thankful to them. I would also like to thank my academic director, \textbf{Prof. Claire David}, and the head of the AI for Science program tutors, \textbf{Emmanuel Ahenkan}, for their valuable advice, time management recommendations, presence, and support during the writing of this thesis and throughout the academic year. To the team at \textbf{Google DeepMind}, without whom this dream would not have been possible, I express my sincere gratitude for their financial support and for the arrangements made for this program. Finally, to everyone who contributed in one way or another to the completion of this work and to the success of this academic year, I extend my deepest thanks.

\newpage

\bibliographystyle{unsrt}
\bibliography{biblio_jmm}

\end{document}